\documentclass{lmcs}
\pdfoutput=1

\usepackage{lastpage}
\lmcsdoi{20}{4}{18}
\lmcsheading{}{\pageref{LastPage}}{}{}%
{Apr.~28,~2022}{Dec.~03,~2024}{}
\usepackage[utf8]{inputenc}


\usepackage{hyperref}

\graphicspath{{./figures/}}

\title{Fine-grained Meta-Theorems for Vertex Integrity} 

\thanks{Partially supported by ANR JCJC projects ``ASSK'' (ANR-18-CE40-0025-01) and ``S-EX-AP-PE-AL'' (ANR-21-CE48-0022)}

\author{Michael Lampis\lmcsorcid{0000-0002-5791-0887}}[a]
\author{Valia Mitsou}[b]

\address{Université Paris-Dauphine, PSL University, CNRS, LAMSADE, 75016, Paris, France}
\email{michail.lampis@lamsade.dauphine.fr}

\address{IRIF, Université Paris-Cité, CNRS, 75205, Paris, France}
\email{vmitsou@irif.fr}

\newcommand{\myc}{\ensuremath\mathcal{C}}

\begin{document}

\begin{abstract}

Vertex Integrity is a graph measure which sits squarely between two more
well-studied notions, namely vertex cover and tree-depth, and that has recently
gained attention as a structural graph parameter. In this paper we investigate
the algorithmic trade-offs involved with this parameter from the point of view
of algorithmic meta-theorems for First-Order (FO) and Monadic Second Order
(MSO) logic.  Our positive results are the following: (i) given a graph $G$ of
vertex integrity $k$ and an FO formula $\phi$ with $q$ quantifiers, deciding if
$G$ satisfies $\phi$ can be done in time $2^{O(k^2q+q\log q)}+n^{O(1)}$; (ii)
for MSO formulas with $q$ quantifiers, the same can be done in time
$2^{2^{O(k^2+kq)}}+n^{O(1)}$. Both results are obtained using kernelization
arguments, which pre-process the input to sizes $2^{O(k^2)}q$ and
$2^{O(k^2+kq)}$ respectively. 

The complexities of our meta-theorems are significantly better than the
corresponding meta-theorems for tree-depth, which involve towers of
exponentials. However, they are worse than the roughly $2^{O(kq)}$ and
$2^{2^{O(k+q)}}$ complexities known for corresponding meta-theorems for vertex
cover. To explain this deterioration we present two formula constructions which
lead to fine-grained complexity lower bounds and establish that the dependence
of our meta-theorems on $k$ is the best possible.  More precisely, we show that it
is not possible to decide FO formulas with $q$ quantifiers in time
$2^{o(k^2q)}$, and that there exists an MSO formula which cannot
be decided in time $2^{2^{o(k^2)}}$, both under the ETH.  Hence, the quadratic
blow-up in the dependence on $k$ is unavoidable and vertex integrity has a
complexity for FO and MSO logic which is truly intermediate between vertex
cover and tree-depth.

\end{abstract}

\maketitle

\section{Introduction}

An algorithmic meta-theorem is a general statement proving that a large class
of problems is tractable. Such results are of great importance because they
allow one to quickly classify the complexity of a new problem, before
endeavoring to design a fine-tuned algorithm. In the domain of parameterized
complexity theory for graph problems, possibly the most well-studied type of
meta-theorems are those where the class of problems in question is defined
using a language of formal logic, typically a variant of First-Order (FO) or
Monadic Second-Order (MSO) logic, which are the logics that allow
quantification over vertices or sets of vertices respectively\footnote{Note
that the version of MSO logic we use in this paper is sometimes also referred
to as MSO$_1$ to distinguish from the version that also allows quantification
over sets of edges.}.  In this area, the most celebrated result is Courcelle's
theorem \cite{Courcelle90}, which states that all properties expressible in MSO
logic are solvable in linear time, parameterized by treewidth and the size of
the MSO formula. In the thirty years since the appearance of this fundamental
result, numerous other meta-theorems in this spirit have followed (we give an
overview of some such results below).

Despite its great success, Courcelle's theorem suffers from one significant
weakness: the algorithm it guarantees for deciding an MSO formula $\phi$ on a
graph $G$ with $n$ vertices and treewidth $k$ has running time $f(k,\phi)\cdot
n$, where $f$ is, in the worst case, a tower of exponentials whose height can
only be bounded as a function of $\phi$. Unfortunately, it has been known since
the work of Frick and Grohe \cite{FrickG04} that this terrible parameter dependence
cannot be avoided, even if one only considers FO logic on trees (or MSO logic
on paths \cite{Lampis14}). This has motivated the study of the complexity of FO and
MSO logic with parameters which are more restrictive than treewidth. In the
context of such parameters, fixed-parameter tractability for all
MSO-expressible problems is already given by Courcelle's theorem, so the goal
is to obtain more ``fine-grained'' meta-theorems which achieve a better dependence on
$\phi$ and $k$.

The two results from this line of research which are most relevant to our paper
are the meta-theorems for vertex cover given in \cite{Lampis12}, and the
meta-theorem for tree-depth given by Gajarsk\'{y} and Hlin{\v{e}}n{\'{y}}
\cite{Gajarsky15}. Regarding vertex cover, it was shown in \cite{Lampis12} that
FO and MSO formulas with $q$ quantifiers can be decided on graphs with vertex
cover $k$ in time roughly $2^{O(kq+q\log q)}$ and $2^{2^{O(k+q)}}$ respectively. Both
of these results were shown to be tight, in the sense that improving their
dependence on $k$ would violate the Exponential Time Hypothesis (ETH). For
tree-depth, it was shown in \cite{Gajarsky15} that FO and MSO formulas with $q$
quantifiers can be decided on graphs with tree-depth $k$ with a complexity that
is roughly $k$-fold exponential. Hence, for fixed $k$, the complexity we obtain
is elementary, but the height of the tower of exponentials increases
with $k$, and this cannot be avoided under the ETH \cite{Lampis14}.

Vertex cover and tree-depth are among the most well-studied measures in parameterized
complexity. In all graphs $G$ we have
$\textrm{vc}(G)+1\ge\textrm{td}(G)\ge\textrm{pw}(G)\ge\textrm{tw}(G)$, so these
parameters form a natural hierarchy with pathwidth and treewidth, with vertex
cover being the most restrictive. As explained above, the distance between the
performance of meta-theorems for vertex cover (which are double-exponential for
MSO) and for tree-depth (which give a tower of exponentials of height
$\textrm{td}$) is huge, but conceptually this is perhaps not surprising.
Indeed, one could argue that the structural distance between graphs of vertex
cover $k$ from the class of graphs of tree-depth $k$ is also huge. As a
reminder, a graph has vertex cover $k$ if we can delete $k$ vertices to obtain
an independent set; while a graph has tree-depth $k$ if there exists $k'\le k$
such that we can delete $k'$ vertices to obtain a disjoint union of graphs of
tree-depth $k-k'$. Clearly, the latter (inductive) definition is more powerful
and covers vastly more graphs, so it is natural that model-checking should be
significantly harder for tree-depth. 

The landscape of parameters described above indicates that there should be
space to investigate interesting structural parameters \emph{between} vertex
cover and tree-depth, exactly because the distance between these two is large
in terms of generality and complexity. One notion that has recently attracted
attention in this area is \emph{Vertex Integrity} \cite{DrangeDH16}, denoted as
$\iota(G)$.  A graph has vertex integrity $k$ if there exists $k'\le k$ such
that we can delete $k'$ vertices and obtain a disjoint union of graphs of
\emph{size} at most $k-k'$. Hence, the definition of vertex integrity is the
same as for tree-depth, except that we replace the inductive step by simply
bounding the size of the components that result after deleting a separator of
the graph.  This produces a notion that is more restrictive than tree-depth,
but still significantly more general than vertex cover (where the resulting
components must be singletons). In all graphs $G$, we have
$\textrm{vc}(G)+1\ge\iota(G)\ge\textrm{td}(G)$, so it becomes an interesting
question to investigate the complexity trade-off associated with these
parameters, that is, how the complexity of various problems deteriorates as we
move from vertex cover, to vertex integrity, to tree-depth. This type of study
was recently undertaken systematically for many problems by Gima et al.\
\cite{Gima21}. In this paper we make an investigation in the same direction
from the lens of algorithmic meta-theorems.

\subparagraph*{Our results} We consider the problem of verifying whether a
graph $G$ satisfies a property given by an FO or MSO formula with $q$
quantifiers, assuming $\iota(G)\le k$. Our goal is to give a fine-grained
determination of the complexity of this problem as a function of $k$. We obtain
the following two positive results:

\begin{enumerate}

\item FO formulas with $q$ quantifiers can be decided in time $2^{O(k^2q+q\log
q)}+n^{O(1)}$.

\item MSO formulas with $q$ vertex and set quantifiers can be decided in time
$2^{2^{O(k^2+kq)}}+n^{O(1)}$.

\end{enumerate}

Hence, we obtain meta-theorems stating that any problem that can be expressed
in FO or MSO logic can be solved in the aforementioned times. Both of these
results are obtained through a kernelization argument, similar in spirit to the
arguments used in the meta-theorems of \cite{Gajarsky15,Lampis12}. To describe the main
idea, recall that if $\iota(G)\le k$, then there exists a separator $S$ of size
at most $k$, such that removing it will disconnect the graph into components of
size at most $k$. The key now is that these components can be partitioned into
$2^{k^2}$ equivalence \emph{types}, where components of the same type are
isomorphic. We then argue that if we have a large number of isomorphic
components, it is always safe to delete any one of them from the graph, as this
does not change whether the given formula holds (Lemmas \ref{lem:FOmanyComp}
and \ref{lem:MSOmanyComp}). We then complete the argument by applying the
standard brute-force algorithms for FO and MSO logic on the kernels. We note
that, even though we do not expend much effort to optimize the $n^{O(1)}$ terms
in the above results, the hidden exponent is rather reasonable as our
kernelization algorithms can easily be executed in time $n^2$.

We complement the results above by showing that the approach of kernelizing and
then executing the brute-force algorithm is essentially optimal. More
precisely, we show that, under the ETH, it is not possible to obtain a
model-checking algorithm for FO logic running in time $2^{o(k^2q)}n^{O(1)}$;
while for MSO we construct a single formula which cannot be model-checked in
time $2^{2^{o(k^2)}}$.  Hence, the quadratic dependence on $k$, which
distinguishes our meta-theorems from the corresponding meta-theorems for vertex
cover, cannot be avoided. 

\subparagraph*{Related work} The study of structural parameters which trade off
the generality of treewidth for improved algorithmic properties is by now a
standard topic in parameterized complexity. The most common type of work here
is to consider a problem that is intractable parameterized by treewidth and see
whether it becomes tractable parameterized by vertex cover or
tree-depth \cite{BelmonteLM20,DellKLMM17,DvorakK18,FellowsFLRSST11,FialaGK11,GutinJW16,HarutyunyanLM21,KatsikarelisLP18,KatsikarelisLP19,KellerhalsK20,LampisM17,Lampis21}. See \cite{Belmonte0LMO20} for a survey of results
of this type. In this context, vertex integrity has only recently started being
studied as an intermediate parameter between vertex cover and tree-depth, and
it has been discovered that fixed-parameter tractability for several problems
which are W-hard by tree-depth can be extended from vertex cover to vertex
integrity \cite{BodlaenderHKKOO20,DvorakEGKO17,GanianKO21,GanianOR21,Gima21}.
Note that some works use a measure called \emph{core fracture} number, which is
a similar notion to vertex integrity.

Algorithmic meta-theorems are a well-studied topic in parameterized complexity (see \cite{Grohe11} for a survey).
Courcelle's theorem has been extended to the more general notion of
clique-width \cite{CourcelleMR00}, and more efficient versions of these
meta-theorems have been given for the more restricted parameters twin-cover
\cite{Ganian15}, shrub-depth \cite{GanianHNOM19,GanianHNOMR12}, neighborhood
diversity and max-leaf number \cite{Lampis12}. Meta-theorems have also been
given for even more general graph parameters, such as
\cite{Bonnet0TW20,DvorakKT13,FrickG01,Frick04}, and for logics other than FO
and MSO, with the goal of either targeting a wider class of problems
\cite{GanianO13,KnopKMT19,KnopMT19,Szeider11}, or achieving better complexity
\cite{Pilipczuk11}.  Meta-theorems have also been given in the context of
kernelization \cite{BodlaenderFLPST16,EibenGS18,GanianSS16} and approximation
\cite{DawarGKS06}. To the best of our knowledge, the complexity of FO and MSO
model checking parameterized by vertex integrity has not been explicitly
studied before, but since vertex integrity is a restriction of tree-depth and a
generalization of vertex cover, the algorithms of \cite{Gajarsky15} and the
lower bounds of \cite{Lampis12} apply in this case.

\section{Definitions and Preliminaries}

First, let us formally define the notion of vertex integrity of a graph.

\begin{defi} For a graph $G$, we define its vertex integrity $\iota(G)$
as the minimal value that satisfies the following: there exists a set
$S\subseteq V(G)$ such that, if $S'\subseteq V(G)$ is the set of vertices of
the largest connected component of $G\setminus S$ then $|S| + |S'| \le
\iota(G)$.  \end{defi}

Note that in the definition above, the separator $S$ is not necessarily a
minimum-sized (or even minimal) separator of $G$. For example, if we take two
stars $K_{1,n}$ and connect their centers, the resulting graph $G$ has
$\iota(G)=3$, as witnessed by the set $S$ that contains both centers; however,
the set $S$ is not a minimal separator of the graph, as either center alone is
also a separator.

Drange et al.\ \cite{DrangeDH16} have shown that deciding if a graph has
$\iota(G)\le k$ admits a kernel of order $O(k^3)$. More strongly, their
kernelization algorithm allows one to obtain an optimal separator $S$ for the
original instance from an optimal separator of the reduced instance. Hence,
given a graph $G$ that is promised to have vertex integrity $k$, we can execute
this kernelization algorithm and then look for the optimal separator $S$ in the
kernel. As a result, finding a separator $S$ proving that $\iota(G)\le k$ can
be done in time roughly $k^{3k}+n^2$, where the latter term comes from the
running time of the kernelization algorithm of \cite{DrangeDH16} and the former
represents all possible choices of $k$ vertices from a graph of order $k^3$.
Since this running time is dominated by the running times of our meta-theorems,
we will always silently assume that the separator $S$ is given in the input
when the input graph has vertex integrity $k$.

A main question that will interest us is whether a graph satisfies a property
expressible in First-Order (FO) or Monadic Second-Order (MSO) logic. Let us
briefly recall the definitions of these logics. We use $x_i, i\in\mathbb{N}$ to
denote vertex (FO) variables and $X_i, i\in \mathbb{N}$ to denote set (MSO)
variables. Vertex variables take values from a set of vertex constants $U =
\{u_i, i\in \mathbb{N}\}$, whereas vertex set variables take values from a set
of vertex set constants $D = \{D_i, i\in \mathbb{N}\}$.

Now, given a graph $G$, in order to say that the assignment of a vertex variable $x_i$ or a vertex set variable $X_i$ to a constant corresponds to a particular vertex or vertex set of $G$, we make use of a \emph{labeling} function $\ell$ that maps vertex constants to vertices of $V(G)$ and of a \emph{coloring} function $\myc$ that maps vertex set constants to vertex sets of $V(G)$. More formally, $\ell, \myc$ are partial functions $\ell:U \rightarrow V(G)$ and $\myc: D \rightarrow 2^{V(G)}$. The functions may be undefined for some constants, for example, if $\ell$ is not defined for the constant $u_i$ we write $\ell(u_i) \uparrow$.

\begin{defi} Suppose we are given a triplet $G,\ell,\myc$, a vertex $v\in
V(G)$ is said to be \emph{unlabeled} if there does not exist $u_i\in U$ such
that $\ell(u_i) = v$.  A set of vertices $C_1\subseteq V(G)$ is unlabeled if
all the vertices of $C_1$ are unlabeled.  \end{defi}

\begin{defi} We say that two labeling functions $\ell, \ell'$
\emph{agree} on a constant $u_i$ if either they are both undefined on $u_i$ or
$\ell(u_i) = \ell'(u_i)$. Similarly,  two coloring functions $\myc, \myc'$
agree on $D_i$ if they are both undefined or $\myc(D_i)=\myc'(D_i)$.
\end{defi}

\begin{defi} Suppose we are given two triplets $G_1, \ell_1, \myc_1$ and
$G_2, \ell_2, \myc_2$ and a bijective function $f:V(G_1) \to V(G_2)$. For
$C_1\subseteq V(G_1)$, we define $f(C_1) = \bigcup_{v\in C_1}\{f(v)\}$. We say
that $V(G_1)$ and $V(G_2)$ \emph{have the same labelings for $f$} if for all
$u_i\in U$, either both $\ell_1(u_i), \ell_2(u_i)$ are undefined or
$f(\ell_1(u_i)) = \ell_2(u_i)$; we say that $V(G_1)$ and $V(G_2)$ \emph{have
the same colorings for $f$} if for all $D_i\in D$, either both $\myc_1(D_i),
\myc_2(D_i)$ are undefined or $f(\myc_1(D_i)) = \myc_2(D_i)$.  \end{defi}

\begin{defi}An \emph{isomorphism} between two triplets $G_1,\ell_1,\myc_1$ and
$G_2,\ell_2,\myc_2$ is a bijective function
$f:V(G_1)\to V(G_2)$ such that (i) for all $v,w \in V(G_1)$ we have $(v,w)\in
E(G_1)$ if and only if $(f(v),f(w))\in E(G_2)$, (ii)  $V(G_1)$ and $V(G_2)$ have the same labelings and colorings for $f$. Two triplets $G_1,\ell_1,\myc_1$ and
$G_2,\ell_2,\myc_2$ are \emph{isomorphic} if there exists an isomorphism between them.
\end{defi}

\begin{defi}\label{def:type} Suppose we are given a triplet
$G,\ell,\myc$.  We say that two sets $C_1\subseteq V(G)$ and $C_2\subseteq
V(G)$ \emph{have the same type} if there exists an isomorphism $f : V(G) \to
V(G)$ between the triplets $G,\ell,\myc$ and $G,\ell,\myc$ such that $f$ maps
elements of $C_1$ to $C_2$ and vice versa and elements from $V(G)\setminus
(C_1\cup C_2)$ to themselves.  \end{defi}

Notice that only for vertices that do not belong in the sets $C_1$ and $C_2$
(which $f$ maps to themselves) we can have that $f(\ell(u_i)) = \ell(u_i)$.
Indeed, if $C_1$ and $C_2$ are disjoint and a vertex $v\in C_1$ is labeled,
since the isomorphism $f$ would have to map it to a vertex $v'\in C_2$, we
would have $v'\neq v$. But in this case, $f$ would not correctly preserve the
labels between the triplets $G,\ell,\myc$ and $G,\ell,\myc$. This leads to the
following observation:

\begin{obs}\label{obs:unlabeled} In order for two disjoint sets $C_1$
and $C_2$ to have the same type, they should necessarily be unlabeled (that is,
for all $u_i$, we have $\ell(u_i) \not\in C_1\cup C_2$).  \end{obs}

\begin{defi} Suppose we are given a triplet $G,\ell,\myc$ and a set
$C_1\subseteq V(G)$.  The \emph{restriction} of $\myc$ to $G\setminus C_1$ is a
function $\myc':D\to V(G)\setminus C_1$ such that $\myc'(D_i) = \myc(D_i)
\setminus C_1$ for all $D_i\in D$ for which $\myc(D_i)\cap C_1 \neq \emptyset$
and $\myc, \myc'$ agree on the rest of $D$.  \end{defi}

An MSO formula is a formula produced by the following grammar, where $X$ represents a set variable, $x$ a vertex variable, $y$ a vertex variable or vertex constant, and $Y$ a set variable or constant:
\begin{eqnarray*}  
\phi & \to & \exists X. \phi\ |\ \exists x.\phi\ |\ \phi \lor \phi\ |\ \neg \phi\ |\ y\sim y\ |\ y=y\ |\ y\in Y
\end{eqnarray*}

The operations above are vertex set quantification, vertex quantification,
disjunction, negation, edge relation, vertex equality, and set inclusion
respectively.  Their semantics are defined inductively in the usual way: given
a triplet $G,\ell,\myc$ and an MSO formula $\phi$, we say
that the graph satisfies the property described by $\phi$, or simply that
$G,\ell,\myc$ models $\phi$, and write $G,\ell,\myc \models \phi$ according to
the following rules: 

\begin{itemize}

\item $G,\ell,\myc \models u_i \in D_j$ if $\ell(u_i)$ and $\myc(D_j)$ are
defined and $\ell(u_i)\in \myc(D_j)$.

\item $G,\ell,\myc \models u_i = u_j$ if $\ell(u_i),\ell(u_j)$ are defined and
$\ell(u_i)=\ell(u_j)$.

\item $G,\ell,\myc \models u_i \sim u_j$ if $\ell(u_i),\ell(u_j)$ are defined
and $(\ell(u_i),\ell(u_j))\in E(G)$.

\item $G,\ell,\myc \models \phi \lor \psi$ if $G,\ell,\myc \models \phi$ or $G,\ell,
\myc\models \psi$.

\item $G,\ell,\myc \models \neg \phi$ if it is not the case that $G,\ell,\myc \models
\phi$.

\item $G,\ell,\myc \models \exists x_i. \phi$ if there exists $v\in V(G)$ such that
$G,\ell',\myc \models \phi[ x_i \setminus u_i ]$, where $\ell(u_i) \uparrow$, $\phi [ x_i \setminus u_i]$ is the formula obtained from $\phi$ if we replace every free occurrence of $x_i$ with the
(new) constant $u_i$ and $\ell':U \to V(G)$ is a partial function for which $\ell'(u_i) = v$, and $\ell',\ell$ agree on all other values $u_j\neq u_i$.

\item $G,\ell,\myc \models \exists X_i. \phi$ if there exists $S\subseteq V(G)$ such
that $G,\ell,\myc' \models \phi[ X_i \setminus D_i ]$, where $\myc(D_i)\uparrow$, $\phi [ X_i \setminus D_i ]$ is the formula obtained from $\phi$ if we replace every free occurrence of $X_i$ with the (new) constant $D_i$ and $\myc':D\to 2^{V(G)}$ is a partial function for which $\myc'(D_i) = S$ and $\myc', \myc$ agree on all other values $D_j\neq D_i$.

\end{itemize}

If none of the above applies then $G,\ell,\myc$ does not model $\phi$ and we
write $G,\ell,\myc \not\models \phi$. Observe that, from the syntactic rules
presented above, a formula can have free (non-quantified) variables. However,
we will only define model-checking for formulas without free variables (also
called sentences). Slightly abusing notation, we will write $G\models \phi$ to
mean $G,\ell,\myc\models \phi$ for the nowhere defined functions $\ell,\myc$.
Note that our definition does not contain conjunctions or universal
quantifiers, but these can be obtained from disjunctions and existential
quantifiers using negations in the usual way, so we will use them freely when
constructing formulas.

An FO formula is defined as an MSO formula that uses no set variables $X_i$. In
the remainder, we will assume that all formulas are given to us in prenex form,
that is, all quantifiers appear in the beginning of the formula. Recall that it
is a well-known fact that all FO and MSO formulas can be converted to prenex
form without increasing the number of quantifiers, so our restriction is
without loss of generality. We call the problem of deciding whether
$G,\ell,\myc \models \phi$ the model-checking problem.

We recall the following basic fact, which we state without proof (a standard
proof can be obtained by structural induction):

\begin{lem}\label{lem:iso} Let $G_1,\ell_1,\myc_1$ and $G_2,\ell_2,\myc_2$ be
two isomorphic triplets.  Then, for all MSO formulas $\phi$ we have
$G_1,\ell_1,\myc_1\models \phi$ if and only if $G_2,\ell_2,\myc_2\models \phi$.
\end{lem} 

\section{FPT Algorithms for FO and MSO Model-Checking Parameterized by Vertex Integrity} 

In this section we prove Theorems~\ref{thmFOalg} and~\ref{thmMSOalg}. The statements appear right below. 

\begin{thm}\label{thmFOalg} Suppose we are given a graph $G$ with $\iota(G)
\le k$ and an FO formula $\phi$ in prenex form having at most $q$ quantifiers.
Then deciding if $G \models \phi$ can be solved in time $(2^{O(k^2)}\cdot q)^q
+ |G|^{O(1)}$.  \end{thm}

\begin{thm}\label{thmMSOalg} Suppose we are given a graph $G$ with
$\iota(G) \le k$ and an MSO formula $\phi$ in prenex form having at most $q_1$
vertex variable quantifiers and at most $q_2$ vertex set variable quantifiers.
Then deciding if $G \models \phi$ can be solved in time $\left(
2^{2^{O(k^2+kq_2)}}\cdot q_1\right)^{q_1} + |G|^{O(1)} $.  \end{thm}

The proofs are heavily based on Lemmata~\ref{lem:FOmanyComp} and~\ref{lem:MSOmanyComp}. The first, which is about FO Model-Checking, says that if we have at least $q+1$ components of the same type then we can erase one such component from the graph. The reason essentially is that, if $G,\ell,\myc$ models $\phi$ by labeling a vertex $v$ that belongs to the component to be removed, we can replace that vertex by a corresponding vertex in another component having the same type. Notice that the formula has $q$ quantifiers and thus the graph will have $q$ labels after the assignment. Since we have $q+1$ components of the same type, for one of these components the vertex that corresponds to $v$ will be unlabeled.

The second, which is about MSO Model-Checking, says that since we can quantify
over sets of vertices, unlike the case for FO, each set quantification can
potentially affect a large number of components that originally had the same
type (by coloring its intersection with each of them). However, since each
component has size at most $k$, we have $2^k$ ways that the quantified set can
overlap with the components. Thus, if we originally had a sufficiently large
number of same type components, even after the coloring, we will still have a
sufficient number of components that are of the same type, such that even if we
remove one such component the answer of the problem will not change.

Lemmata~\ref{lem:FOmanyComp} and~\ref{lem:MSOmanyComp}, together with the fact
that there exists a bounded number of types of components, give the kernels
(Lemma~\ref{lem:FOkernel} for FO and Lemma~\ref{lem:MSOkernel} for MSO). 

\begin{lem}\label{lem:FOmanyComp} Suppose we are given a triplet $G, \ell,
\myc$ having $q+1$ disjoint vertex sets $C_1, C_2, \ldots, C_{q+1}$ of the same
type and $\phi$ an FO formula in prenex form having $q$ quantifiers. Then $G,
\ell,\myc \models \phi$ if and only if $G\setminus C_1, \ell,\myc'\ \models
\phi$, where $\myc'$ is the restriction of $\myc$ to $V(G) \setminus C_1$.
\end{lem}

\begin{proof}
We proceed by induction on the structure of the formula $\phi$. 
\begin{enumerate}

\item For $\phi := u_i \in D_j$, $\phi := u_1 = u_2$, or $\phi := u_1 \sim
u_2$. From Observation~\ref{obs:unlabeled} the sets are unlabeled. Thus, there
is no $ v \in C_1$ for which $\ell(u_1) = v$ or $\ell(u_2)=v$. Thus the
statement of the lemma holds for the base case.

\item For $\phi := \phi_1 \lor \phi_2$ or $\phi := \neg \phi_1$. From the inductive hypothesis, we have that $G, \ell,\myc \models \phi_1$ if and only if $G\setminus C_1, \ell,\myc'\ \models \phi_1$ and that $G, \ell,\myc \models \phi_2$ if and only if $G\setminus C_1, \ell,\myc'\ \models \phi_2$. It is easy to see that the statement of the lemma holds also for $\phi$.
\item The most interesting case is for $\phi := \exists x_i. \phi'$. 
If $G,\ell, \myc \models \phi$ then from the definition of the semantics of $\phi$ there exists $v\in V(G)$ such that
$G,\ell',\myc \models \phi[ x_i \setminus u_i ]$ with $\ell(u_i) \uparrow$ and $\ell':U \to V(G)$ being a partial function for which $\ell'(u_i) = v$, and $\ell'$ agrees with $\ell$ on all other values $u_j\neq u_i$. 

First we prove that without loss of generality $v\not\in C_1$. Suppose that
$v\in C_1$. Since $C_1$ and $C_2$ have the same type on $G,\ell,\myc$, by
\autoref{def:type} there exists an isomorphism $f: C_1 \to C_2$. Consider now a
labeling function $\ell'': U \to V(G)$ where $\ell''(u_i) = f(\ell'(u_i)) =
f(v)$, otherwise $\ell', \ell''$ agree on $u_j \neq u_i$. Observe that
$G,\ell',\myc$ and $G,\ell'',\myc$ are isomorphic, thus from
Lemma~\ref{lem:iso} we have that $G,\ell',\myc \models \phi$ if and only if
$G,\ell'',\myc \models \phi$. In that case, instead of $v\in C_1$ we shall
consider $f(v) \in C_2$. Thus, from now on we can assume that $v\not\in C_1$ 

For the triplet $G,\ell',\myc$ we have that $q$ of the sets $C_1, C_2, \ldots,
C_{q+1}$ are still unlabeled and have the same type ($C_1$ is among them). Also
$\phi'$ has $q-1$ quantifiers. Thus, by the inductive step, $G,\ell',\myc
\models \phi'$ if and only if $G\setminus C_1, \ell', \myc' \models \phi'$.
Since $v\in V(G)\setminus C_1$, we have that $G\setminus C_1,\ell, \myc'
\models \phi$.

For the other direction, observe that $v\in V(G)\setminus C_1$ implies that
$v\in V(G)$. Thus the statement holds with similar reasoning as above. \qedhere
\end{enumerate}\end{proof}

Note that \autoref{lem:FOmanyComp} can be seen as a kind of ``pumping lemma'', as
it states that, after a certain point, adding components of the same type to a
graph does not affect whether the graph satisfies a formula.

\begin{lem}\label{lem:FOkernel} For a triplet $G, \ell, \myc$ with vertex
integrity $\iota(G)\le k$ and with $\ell, \myc$ everywhere undefined and for a
formula $\phi$ with $q$ quantifiers, \textsc{FO Model Checking} has a kernel of
size $O(2^{k^2} \cdot q \cdot k)$, assuming we are given in the input
$S\subseteq V(G)$ such that the largest component of $G\setminus S$ has size at
most $k-|S|$.  \end{lem}

\begin{proof} We give a polynomial-time algorithm to calculate an upper bound
on the number of components of $G\setminus S$ having the same type. Observe
that types are only specified by the neighborhoods of the vertices of the
components ($\ell$ and $\myc$ are everywhere undefined thus there are no labels
or colors on $G$). 

First, we arbitrarily number the vertices of $S$ and of each component. In
order to classify the components into types, we map each component $C_i$ to a
vector $[N_1, N_2, \ldots, N_{|C_i|}]$, where $N_j$ is an ordered set
containing the (numbered) neighbors of the $j^{\textrm{th}}$ vertex of $C_i$
(starting from the neighbors in $S$). Clearly, if two components have the same
vectors, then they also have the same type, as witnessed by the isomorphism
that maps the $i$-th vertex of one to the $i$-th vertex of the other.

Since each component has at most $k$ vertices and each vertex has at most $2^k$
different types of neighborhoods $N_j$, we can have at most $2^{k^2}$ vectors,
thus at most $2^{k^2}$ types of components.  Furthermore, since we are given
$S$, we can test in polynomial time if two components have the same type under
the arbitrary numbering we used.  From Lemma~\ref{lem:FOmanyComp}, if more than
$q$ components have the same type we can remove one such component without
changing the answer of the problem, thus we can in polynomial time either
reduce the graph or conclude that each component type appears at most $q$
times.  In the end we will have at most $2^{k^2} \cdot q$ components, each
having at most $k$ vertices, thus the result.  \end{proof}

By applying the straightforward algorithm which runs in time ${|V(G)|}^q \cdot poly(|G|)$ for \textsc{FO Model Checking}, together with Lemma~\ref{lem:FOkernel} we get the complexity promised by Theorem~\ref{thmFOalg}.

In order to prove Theorem~\ref{thmMSOalg} we need a stronger version of Lemma~\ref{lem:FOmanyComp}.

\begin{lem}\label{lem:MSOmanyComp} Suppose we are given a triplet $G, \ell,
\myc$ with at least $q'= 2^{k\cdot q_2} \cdot q_1 + 1 $ disjoint vertex sets
$C_1, C_2, \ldots, C_{q'}$ having the same type and sizes at most $k$ and an
MSO formula $\phi$ in prenex form with $q_1$ many FO quantifiers and $q_2$ many
MSO quantifiers. Then $G,\ell,\myc \models \phi$ if and only if $G\setminus
C_1, \ell,\myc_1 \models \phi$, where $\myc_1$ is the restriction of $\myc$ to
$V(G)\setminus C_1$.  \end{lem}

\begin{proof}

We proceed by induction on the structure of $\phi$. We can reuse the arguments
of Lemma~\ref{lem:FOmanyComp},  except for the case where $\phi := \exists X_i.
\phi'$, so we focus on this case.

For the one direction, if $G,\ell,\myc \models \phi$, from the definition of the semantics of $\phi$, then there exists $S\subseteq V(G)$ such that $G,\ell,\myc' \models \phi[ X_i \setminus D_i ]$ with $\myc(D_i) \uparrow$ and $\myc':D \to 2^{V(G)}$ being a partial function for which $\myc'(D_i) = S$, and $\myc'$ agrees with $\myc$ on all other values $D_j\neq D_i$. 

Since each of the vertex sets $C_1, C_2, \ldots, C_{q'}$ has size at most $k$,
there are at most $2^k$ possible ways for $S$ to intersect with each of them.
Therefore, by pigeonhole principle, one such intersection appears in at least
$\lceil\frac{q'}{2^k}\rceil=2^{k(q_2-1)}\cdot q_1 +1$ sets, call that group
$M$.  In order to be able to apply the inductive hypothesis, we need to prove
that, without loss of generality, $C_1\in M$. 

Suppose that $C_1 \not \in M$. We will do a ``swapping'' of $C_1$ with a vertex set (say $C_2$ without loss of generality) that does belong in the group $M$. Since $C_1$ and $C_2$ have the same type, that means that there exists an isomorphism $f : C_1 \to C_2$. 

We consider a new coloring function $\myc''$ that agrees with $\myc'$ everywhere but on the constant $D_i$. This new coloring function will map $D_i$ to the set of vertices $S'$ (instead of $S$), where we have replaced every $v\in S \cap C_1$ with $f(v)$ and every $v\in S\cap C_2$ with $f^{-1}(v)$ (see Figure~\ref{fig:col}). More formally, $\myc''(D_i) = S'$ where $S' = (S\setminus (C_1\cup C_2)) \cup f(C_1 \cap S) \cup f^{-1}(C_2 \cap S)$. Then the triplets $G, \ell, \myc'$ and $G, \ell, \myc''$ are isomorphic and from Lemma~\ref{lem:iso} we have that $G, \ell, \myc' \models \phi$ iff $G, \ell, \myc'' \models \phi$. From now on we assume that $C_1$ belongs in $M$.

\begin{figure}
\centering
\includegraphics[scale=0.5]{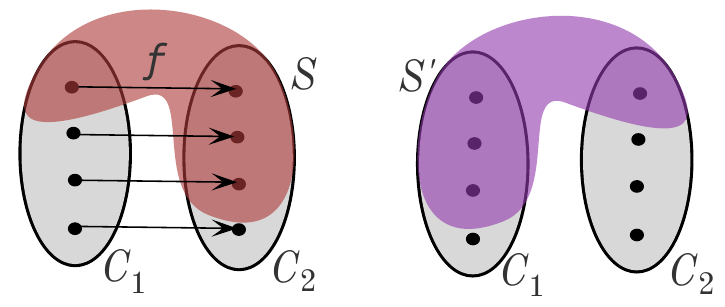}
\caption{The way the vertex set $S'$ intersects the vertex sets $C_1$ and $C_2$.}
\label{fig:col}
\end{figure}

For the triplet $G, \ell, \myc'$, the sets in $M$ have all the same type and
$|M|\ge 2^{k(q_2-1)}\cdot q_1 +1$. Furthermore, the formula $\phi'$ has $q_1$
many FO and $q_2-1$ many MSO quantifiers. Therefore, by the inductive
hypothesis we can remove a set from $M$ and the answer of the problem will not
change, in other words we have that $G, \ell, \myc' \models \phi'$ if and only
if $G\setminus C_1, \ell, \myc'_1 \models \phi'$, where $\myc'_1$ is the
restriction of $\myc'$ on $V(G)\setminus C_1$. From the semantics of $\phi$ we
have that $G\setminus C_1, \ell, \myc_1 \models \phi.$

For the other direction, if $G\setminus C_1,\ell,\myc_1 \models \phi$ then there exists $S_1\subseteq V(G)\setminus C_1$ such that $G\setminus C_1,\ell,\myc_1' \models \phi[ X_i \setminus D_i ]$ with $\myc_1(D_i) \uparrow$ and $\myc_1$ being a partial coloring function for which $\myc_1'(D_i) = S_1$, and $\myc_1'$ agrees with $\myc_1$ on all other values $D_j\neq D_i$. 

As previously, $S_1$ partitions $C_2,\ldots,C_{q'}$ into $2^k$ equivalence
classes, depending on the intersection of each set with $S_1$, such that sets
placed in the same class (i.e. having isomorphic intersection with $S_1$) have
the same type in $G\setminus C_1,\ell,\myc_1'$.  Hence, one of these classes
has size at least $\frac{q'-1}{2^k} = 2^{k(q_2-1)}\cdot q_1$, call this class
$M'$.  We construct a triplet $G,\ell,\myc^*$ as follows: let $C_j\in M'$ and
$f'$ be the isomorphism from $C_j$ to $C_1$. We set that $\myc^*$ agrees with
$\myc$ on all sets except $D_i$; and for $D_i$ we have $\myc^*(D_i) =
\myc_1'(D_i) \cup f'(S_1\cap C_j)$. In other words, we define $\myc^*$ in such
a way that the set $C_1$ has the same type as all sets of the class $M'$. But
then we have $|M'\cup\{C_1\}|\ge 2^{k(q_2-1)}\cdot q_1+1$ sets of the same type
and by inductive hypothesis we have $G,\ell,\myc^*\models \phi[ X_i \setminus
D_i]$.  Therefore, by the semantics of MSO we have $G,\ell,\myc\models \phi$.
\end{proof}

\begin{lem}\label{lem:MSOkernel} For a triplet $G, \ell, \myc$ with vertex
integrity $\iota(G)\le k$ and with $\ell, \myc$ everywhere undefined and for a
formula $\phi$ with $q_1$ many FO quantifiers and $q_2$ many MSO quantifiers,
\textsc{MSO Model Checking} has a kernel of size $O(2^{(k^2+kq_2)} \cdot q_1
\cdot k)$, assuming we are given in the input $S\subseteq V(G)$ such that the
largest component of $G\setminus S$ has size at most $k-|S|$.  \end{lem}

\begin{proof}
The proof is the same as for Lemma~\ref{lem:FOkernel}. The only thing that changes is the number of same-type components required to have before removing one such component ($q'$ required by Lemma~\ref{lem:MSOmanyComp} versus $q+1$ required by Lemma~\ref{lem:FOmanyComp}). 
\end{proof}

Applying the straightforward algorithm for MSO Model-Checking that runs in $2^{q_2\cdot V(G)} \cdot V(G) ^{q_1} \cdot poly|G|$ and Lemma~\ref{lem:MSOkernel} gives the complexity promised by Theorem~\ref{thmMSOalg}.

\section{Lower Bounds}

In this section we show that the dependence of our meta-theorems on vertex
integrity cannot be significantly improved, unless the ETH is false. Our
strategy will be to present a unified construction which, starting from an
arbitrary graph $G$ with $n$ vertices, produces a new graph $H(G)$, with
small vertex integrity, such that we can deduce if two vertices of $G$ are
connected using appropriate FO formulas that describe properties of $H$.  This
will, in principle, allow us to express an FO or MSO-expressible property of
$G$ as a corresponding property of $H(G)$, and hence, if the original property
is hard, to obtain a lower bound on model-checking on $H$.  Let us describe
this construction in more details.

\subparagraph*{Construction} We are given a graph $G$ on $n$ vertices,
say $V(G)=\{v_1,\ldots, v_{n}\}$, and $m$ edges. Let $k=\lceil \sqrt{\log n}\
\rceil$.  We construct a graph $H$ as follows:

\begin{enumerate}

\item We begin constructing $V(H)$ by forming $n+m+1$ sets of vertices, called
$S$, $W_1,\ldots,W_n$, and $Y_1,\ldots,Y_m$. We have $|S|=2k$, $|W_i|=k$ for
all $i\in [n]$, and $|Y_j|=2k+1$ for all $j\in [m]$. The vertices of $S$ are
numbered arbitrarily as $s_1,s_2,\ldots,s_{2k}$.

\item Internally, $S$ induces an independent set, each $W_i$, for $i\in [n]$
induces a clique, and each $Y_j$, for $j\in [m]$ induces a graph made up of two
disjoint cliques of size $k$, denoted $Y_j^1, Y_j^2$, and a vertex connected to
all $2k$ vertices of the cliques $Y_j^1, Y_j^2$.

\item For each $i\in [n]$, we attach a leaf to each vertex of $W_i$. For each
$j\in [m]$, we attach two leaves to each vertex of $Y_j^1$, three leaves to
each vertex of $Y_j^2$, and four leaves to the remaining vertex of $Y_j$.

\item For each $i\in [n]$, number the vertices of $W_i$ arbitrarily as
$w_{(i,1)}, w_{(i,2)},\ldots, w_{(i,k)}$. For each $\beta\in [k]$ we connect
$w_{(i,\beta)}$ to $s_{\beta}$. Furthermore, let $b_1b_2\ldots b_{k^2}$ be the
binary representation of $i-1$ with the least significant digit first, that is,
a sequence of bits such that $\sum_{\beta} b_{\beta}2^{\beta-1} = i-1$.  Note
that $k^2\ge \log n$, therefore $k^2$ bits are sufficient to represent all
numbers from $0$ to $n-1$.  We partition this binary representation into $k$
blocks of $k$ bits.  For $\beta\in [k]$ we consider the bits
$b_{(\beta-1)k+1}\ldots b_{\beta k}$ and we use these bits to determine the
connections between $w_{(i,\beta)}$ and the vertices $s_{k+1},\ldots, s_{2k}$.
More precisely, for $\beta,\gamma\in [k]$, we set that $w_{(i,\beta)}$ is
connected to $s_{k+\gamma}$ if and only if $b_{(\beta-1)k+\gamma}$ is equal to
$1$.

\item For each $j\in [m]$ we do the following. Suppose the $j$-th edge of $G$
has endpoints $v_{i_1},v_{i_2}$. We number the vertices of $Y_j^1$ as
$y_{(j,1)}^1,\ldots,y_{(j,k)}^1$, and the vertices of $Y_j^2$ as
$y_{(j,1)}^2,\ldots,y_{(j,k)}^2$ in some arbitrary way. Now for all $\beta\in
[k]$ we set that $y_{(j,\beta)}^1$ has the same neighbors in $S$ as
$w_{(i_1,\beta)}$ and $y_{(j,\beta)}^2$ has the same neighbors in $S$ as
$w_{(i_2,\beta)}$.

\end{enumerate}

\begin{figure}
\centering
\includegraphics[scale=0.5]{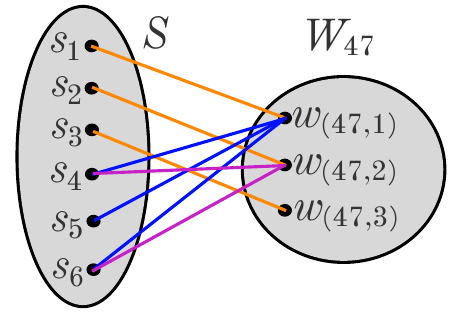}
\caption{The connection between $S$ and the set $W_{47}$. For this example
$k=3$, we can represent up to $2^9$ numbers in binary. In order to represent
$47_{10} = 000101111_2$, we shall connect $w_{(47,1)}$ with $s_4, s_5$ and
$s_6$ in order to represent the three least significant bits (which are all 1),
and $w_{(47,2)}$ with $s_4$ and $s_6$ to represent the next triad of bits. The
three most significant bits are all 0, therefore $w_{(47,3)}$ is not connected
to any of $s_4,s_5,s_6$.} \label{fig:W47} \end{figure}

The construction of our graph is now complete. The intuition behind this
construction is that each clique $W_i$ represents a vertex $v_i\in V(G)$.  In
order to distinguish the vertices, we use the $k^2\ge \log n$ possible edges
between vertices in $W_i$ and the second part of $S$, that is
$\{s_{k+1},\ldots,s_{2k}\}$.  These edges should represent the binary
representation of $i$. See Figure~\ref{fig:W47} for an example. 

Vertices of $H$ may be (arbitrarily) labeled for the purpose of the
construction but for the purpose of Model-Checking the graph $H$ is unlabeled.
In order to give a numbering to the vertices of $W_i$, we use the matching
between $W_i$ and the first $k$ vertices of the set $S$ (the first vertex of
$W_i$ connects to the first vertex of $S$, etc).

The sets $Y_j$ represent edges in $G$. If the $j^\textrm{th}$ edge in $E(G)$ is
the edge $(v_{i_1}v_{i_2})$, then $Y_j^1$ should have the same connections with
$S$ as the set $W_{i_1}$ (similarly $Y_j^2$, $W_{i_2}$).  In order to check in
$H$ whether $(v_{i_1},v_{i_2})$ is an edge, we shall check if there exists a
set $Y_j$ such that each vertex of $Y_j^1$ has the same neighborhood in $S$ as
a vertex of $W_{i_1}$ and each vertex of $Y_j^2$ has the same neighborhood in
$S$ as a vertex of $W_{i_2}$. 

It is crucial here that the construction is such that $W_i,W_{i'}$ are distinguishable
for $i\neq i'$ in terms of their neighborhoods in $S$, that is, there always exists $w\in W_i$ 
for which no $w'\in W_{i'}$ has $N(w)\cap S= N(w')\cap S$. We will show that it is not hard to
express this property in FO logic. Furthermore, the leaves we have attached to
various vertices will allow us to distinguish in FO logic whether a vertex
belongs in a set $W_i$, $Y_j^1$, or $Y_j^2$. 

We now establish some basic properties about $H$ and what can be expressed
about its vertices in FO logic:

\begin{lem}\label{lem:reduction}

There exist FO formulas $\phi_W(x_1),\phi_{Y1}(x_1),\phi_{Y2}(x_1),
\phi_S(x_1)$ using one free variable $x_1$ and FO formulas $\phi_{WY}(x_1,x_2),
\phi_{adj}(x_1,x_2)$ using two free variables $x_1,x_2$, such that any graph
$H$ constructed as described above satisfies the following properties, for any
coloring function $\myc$.

\begin{enumerate}

\item We have $\iota(H) = O(\sqrt{\log n})$ and $|V(H)|=O(n^2\sqrt{\log n})$.

\item For each $i,i'\in [n]$ with $i\neq i'$, there exists a vertex $w\in W_i$ such
that for all $w'\in W_{i'}$ we have $N(w)\cap S\neq N(w')\cap S$.

\item $H,\ell,\myc \models \phi_W[x_1\setminus u_1]$ (respectively $H,\ell,\myc
\models \phi_{Y1}[x_1\setminus u_1]$, $H,\ell,\myc \models
\phi_{Y2}[x_1\setminus u_1]$, $H,\ell,\myc \models \phi_{S}[x_1\setminus u_1]$)
if and only if $\ell(u_1)\in W_i$ for some $i\in [n]$ (respectively
$\ell(u_1)\in Y_j^1$, $\ell(u_1)\in Y_j^2$, for some $j\in [m]$, $\ell(u_1)\in
S$).

\item $H,\ell,\myc\models \phi_{WY}[x_1\setminus u_1][x_2\setminus u_2]$ if and
only if $\ell(u_1)\in W_i$ for some $i\in [n]$, $\ell(u_2)\in Y_j^{\alpha}$ for
some $j\in [m], \alpha\in\{1,2\}$, and for all $\beta\in [k]$ we have
$N(w_{(i,\beta)})\cap S = N(y_{(j,\beta)}^{\alpha})\cap S$.

\item $H,\ell,\myc \models \phi_{adj}[x_1\setminus u_1][x_2 \setminus u_2]$ if
and only if $\ell(u_1)\in W_i$ and $\ell(u_2)\in W_{i'}$ for some $i,i' \in
[n]$ such that  $(v_i, v_{i'})\in E(G)$.  

\end{enumerate}

\end{lem}

\begin{proof}

For the first property, we observe that the largest component of $H\setminus S$
has size at most $10\sqrt{\log n}+2$, while $|S|\le 2\sqrt{\log n}+2$.
Furthermore, we have at most $m+n = O(n^2)$ components after removing $S$.

For the second property, since $i\neq i'$, their binary representations differ
in some bit. Let $\beta,\gamma\in[k]$ be such that if $b_1\ldots b_{k^2}$ is
the binary representation of $i-1$ and  $b_1'\ldots b_{k^2}'$ is the binary
representation of $i'-1$, we have $b_{(\beta-1)k+\gamma} \neq
b'_{(\beta-1)k+\gamma}$. But then, exactly one of $w_{(i,\beta)},
w_{(i',\beta)}$ is connected to $s_{k+\gamma}$. Furthermore, $w_{(i,\beta)}$ is
connected to $s_{\beta}$, but the only neighbor of $s_{\beta}$ in $W_{i'}$ is
$w_{(i',\beta)}$. Hence, $w_{(i,\beta)}$ is the claimed vertex.

For the third property, observe that, in $H$, vertices of $S$ have no leaves
attached, vertices of each $X_i$ have one leaf attached, vertices of $Y_j^1$
have two leaves attached, vertices of $Y_j^2$ have three leaves attached, and
the remaining vertices have four leaves attached.  Furthermore, the only
vertices of degree one in the graph are those which we explicitly added as
leaves. Hence, it suffices to be able to express in FO the property ``$x_1$ has
exactly $c$ leaves attached'', where $c\in\{0,1,2,3\}$.  This is not hard to
do. For example, the following formula expresses the property that $x_1$ has at least two
leaves attached to it:

\begin{eqnarray*} \phi_2(x_1) &:=& \exists x_2 \exists x_3 \big( (x_2\sim x_1) \land (x_3\sim x_1)\land (x_2\neq x_3) \land \\
&& \forall x_4 \left((x_4 = x_1) \lor \left(\neg (x_4 \sim x_2) \land \neg (x_4 \sim x_3)
\right) \right) \big) 
\end{eqnarray*}

Using the same ideas we can construct $\phi_c(x_1)$, for
$c\in \{1,2,3,4\}$ and then $\phi_S(x_1):=\neg \phi_1(x_1)$, $\phi_W(x_1) :=
\phi_1(x_1) \land \neg \phi_2(x_1)$, $\phi_{Y1} := \phi_2 (x_1)\land \neg
\phi_3(x_1)$, $\phi_{Y2}(x_1) := \phi_3(x_1) \land \neg \phi_4(x_1)$.

For the fourth property, we set $\phi_{WY}(x_1,x_2) := \phi_{WY1}(x_1,x_2) \lor \phi_{WY2}(x_1,x_2)$,
where we define two formulas $\phi_{WY\alpha}$ depending on whether $\alpha=1$ or $\alpha=2$. We have
\begin{eqnarray*}
\phi_{WY\alpha}(x_1,x_2)&:=& \phi_W(x_1) \land \phi_{Y\alpha}(x_2) \land \forall x_3 \big( 
(\neg \phi_W(x_3)) \lor (\neg (x_3\sim x_1)\land \neg(x_3=x_1)) \lor \\ 
& &\exists x_4\big( \phi_{Y\alpha}(x_4) \land
(x_4\sim x_2\lor x_4=x_2) \land \\ && \forall x_5 \left( \phi_S(x_5) \to (x_5\sim x_3 \leftrightarrow
x_5\sim x_4) \right)\big) \big)
\end{eqnarray*}

What we are saying here is that $\phi_{WY1}[x_1\setminus u_1][x_2\setminus u_2]$ 
is satisfied if $\ell(u_1)\in W_i, \ell(u_2)\in Y_j^1$, for some $i\in[n], j\in[m]$, 
and for every $x_3\in
W_i$ there exists $x_4\in Y_j^1$  such
that $N(x_3)\cap S = N(x_4)\cap S$. Therefore, if this property holds, then $W_i$
and $Y_j^1$ represent the same vertex of $V$ (similarly for $\phi_{WY2}$).  

For the last property, we set 
\begin{eqnarray*}
\phi_{adj}(x_1,x_2)&:=& \phi_W(x_1) \land \phi_W(x_2) \land\\ &&  
\exists x_3 \exists x_4 \big( (\phi_{Y1}(x_3) \land \phi_{Y2}(x_4)) \lor (\phi_{Y1}(x_4) \land \phi_{Y2}(x_3))\big) \land \\ 
& &\phi_{WY}(x_1,x_3) \land \phi_{WY}(x_2,x_4) \land \exists x_5 (\neg \phi_S(x_5) \land x_3\sim x_5 \land x_4 \sim x_5)
\end{eqnarray*}

In other words, $H, \ell, \myc \models \phi_{adj}[x_1\setminus u_1][x_2\setminus u_2]$ 
if (i) $\ell(u_1)\in W_i$ and $\ell(u_2)\in W_{i'}$, for some $i,i'\in [n]$;
(ii) there exist $x_3$ and $x_4$ such that $x_3\in Y_j^1$ and $x_4\in Y_j^2$ for the
same $j$; this is verified because $x_3$ and $x_4$ have a common neighbor $x_5$ that
does not belong in $S$; (iii) $W_i, W_{i'}$ correspond to the same pair
of vertices as the set $Y_j = Y_j^1 \cup Y_j^2$, which means that $(v_i,v_{i'})\in E(G)$.
\end{proof}

We are now ready to prove our lower bounds.

\begin{thm}\label{thm:hard1} If there exists an algorithm which, given an
FO formula $\phi$ with $q$ quantifiers, an integer $k$, and a graph $G$ on $n$
vertices with $\iota(G)=k$, decides whether $G\models \phi$ in time
$2^{o(k^2q)}n^{O(1)}$, then the ETH is false.  \end{thm}

\begin{proof}

We perform a reduction from $q$-\textsc{Clique}. It is well-known that, given a
graph $G$ on $n$ vertices it is not possible to decide if $G$ contains a clique
of size $q$ in time $n^{o(q)}$, unless the ETH is false \cite{CyganFKLMPPS15}.
We construct the graph $H(G)$, as previously described. We then claim that we
can also construct an FO formula $\phi_C$ such that $\phi_C$ contains $O(q)$
quantifiers and $H,\ell,\myc\models \phi_C$ for the nowhere defined functions
$\ell,\myc$ if and only if $G$ has a $q$-clique.  If we show this, then, since
by \autoref{lem:reduction} we have $k=O(\sqrt{\log n})$, and the size of $H$ is
polynomially related to the size of $G$, the stated running time would become
$2^{o(q(\sqrt{\log n})^2)} n^{O(1)} = n^{o(q)}$ and we refute the ETH.  Our
goal is then to define such an FO formula $\phi_C$. We define 

\begin{eqnarray*} 
\phi_C &:=& \exists x_1 \exists x_2\cdots \exists x_q
\bigwedge_{i\in [q]} \phi_W(x_i) \land \bigwedge_{i,i' \in [q], i\neq i'} (x_i
\neq x_{i'}) \land \\ 
& & \forall x_{q+1} \forall x_{q+2} \Big( \bigwedge_{i\in[q]}
\big(\neg(x_{q+1} = x_i)\big) \lor  \bigwedge_{i\in[q]} \big(\neg(x_{q+2} =
x_i)\big) \lor (x_{q+1} = x_{q+2}) \lor \\ & &\ \ \phi_{adj}(x_{q+1},x_{q+2}) \Big).
\end{eqnarray*}

We now claim that by the construction of $H$, we have that $H,\ell,\myc\models
\phi_C$ if and only if $G$ has a clique. If $G$ has a clique
$\{v_{i_1},v_{i_2},\ldots,v_{i_q}\}$, we map $x_1, x_2,\ldots, x_q$ to
arbitrary vertices of $W_{i_1},\ldots, W_{i_q}$. For the next part of the formula,
either $x_{q+1}, x_{q+2}$ correspond to some (different) $x_i, x_{i'}$ or
the formula is true. Last, we claim that $H,\ell',\myc\models \phi_{adj}[x_{q+1}\setminus u_{i}][x_{q+2}\setminus u_{i'}]$,  
where $x_i,x_{i'}$ are substituted by $u_i, u_{i'}$ and $\ell'(u_i)\in W_i, \ell'(u_{i'})\in W_{i'}$.  
Indeed, because we have a clique in $G$, by construction there
exists a $Y_j$ such that each vertex of $Y_j^1$ has the same neighborhood in
$S$ as $W_{i}$ and each vertex of $Y_j^2$ has the same neighborhood in $S$ as
$W_{i'}$ (or the same with the roles of $Y_j^1, Y_j^2$ reversed). Hence,
$\phi_{adj}$ is satisfied. 

For the converse direction, suppose that
$H,\ell,\myc\models \phi_C$ for the nowhere defined labeling function $\ell$. Then there exists a labeling function $\ell'$ that assigns $\ell'(u_1), \ell'(u_2), \ldots, \ell'(u_q)$ to some vertices of $\bigcup_{i\in [n]}W_i$ and is undefined everywhere else such that $\ell'(u_i) \neq \ell'(u_{i'})$ for $i\neq i'$ and $H,\ell',\myc\models \phi_{C'}$ where 

\[\phi_{C'} := \forall x_{q+1} \forall x_{q+2} \bigwedge_{i\in[q]} \big((x_{q+1} \neq u_i)\big) \lor  \bigwedge_{i\in[q]} \big((x_{q+2} \neq u_i)\big) \lor  (x_{q+1} = x_{q+2}) \lor 
\phi_{adj}(x_{q+1},x_{q+2})\]

We extract a multi-set $S$ of $q$ vertices of $G$ as follows: for $\beta\in
[q]$, if $\ell'(u_\beta)\in W_i$, then we add $v_i$ to $S$. We claim that for
any two elements $v_i,v_{i'}$ of $S$ we have $(v_i,v_{i'})\in E$. If we prove
this, then the vertices of $S$ are distinct and form a $q$-clique in $G$.

Since we have universal quantifications for $x_{q+1}, x_{q+2}$, we can define a
new labeling function $\ell''$, with $\ell''(u_{q+1}) = \ell'(u_i)$ and
$\ell''(u_{q+2}) = \ell'(u_{i'})$, for any $i,i'\in[q], i\neq i'$, with
$\ell'', \ell'$ agreeing everywhere else.  Observe that this selection imposes
that $H,\ell'',\myc\models \phi_{adj}[x_{q+1}\setminus u_i][x_{q+2}\setminus
u_{i'}]$ and from property 5 of Lemma~\ref{lem:reduction} we get that
$\ell'(u_i), \ell'(u_{i'})$ belong to two different $W_j, W_{j'}$ that
correspond to the endpoints of an edge of $G$.  \end{proof}

\begin{thm} There is an MSO formula $\phi$ such that we have the following:
if there exists an algorithm which, given a graph $G$ with $n$ vertices and
$\iota(G)=k$, decides whether $G\models \phi$ in time $2^{2^{o(k^2)}}n^{O(1)}$,
then the ETH is false.  \end{thm}

\begin{proof}

Our strategy is similar to that of \autoref{thm:hard1}, except that we will now
reduce from \textsc{3-Coloring}, which is known not to be solvable in
$2^{o(n)}$ on graphs on $n$ vertices, under the ETH \cite{ImpagliazzoPZ01}. We
will produce a formula $\phi_{Col}$ with the property that $H,\ell,\myc\models
\phi_{Col}$ for the nowhere defined functions $\ell,\myc$ if and only if $G$ is
3-colorable. Since $k=O(\sqrt{\log n})$ an algorithm running in
$2^{2^{o(k^2)}}$ would imply a $2^{o(n)}$ algorithm for 3-coloring $G$,
contradicting the ETH. We define \begin{eqnarray*} \phi_{Col} &:=& \exists X_1
\exists X_2 \exists X_3 \forall x_1 \forall x_2 (x_1\in X_1\lor x_1\in X_2\lor
x_1\in X_3) \land \\ &&\bigwedge_{i=1,2,3} \phi_{adj}(x_1, x_2) \to \big(x_1\in
X_i \to \neg (x_2 \in X_i)\big) \end{eqnarray*}

Assume that $G$ has a proper 3-coloring $c:V\to [3]$. Then we define, for
$\alpha \in [2]$ $S_\alpha = \bigcup_{i:c(v_i) = \alpha} W_i$ and $S_3 = V(H)
\setminus (S_1 \cup S_2)$.  Let $\myc'$ be a coloring function such that
$\myc'(D_\alpha) = S_\alpha$ for $\alpha = 1,2,3$ and $\myc'(D_{\alpha'})
\uparrow$ for $\alpha' \not \in [3]$. We claim that $H,\ell, \myc' \models
\phi_{Col}[X_1\setminus D_1][X_2\setminus D_2][X_3\setminus D_3]$. Indeed, for
any labeling function $\ell'$ that defines only $\ell'(u_1)$ and $\ell'(u_2)$
we have (i) $H,\ell',\myc'\models u_1 \in D_1 \lor u_1 \in D_2 \lor u_1 \in
D_3$ (since $\myc'(D_1), \myc'(D_2), \myc'(D_3)$ is a partition of $V(H)$),
(ii) if $H,\ell',\myc' \models \phi_{adj}[x_1\setminus u_1][x_2\setminus u_2]$
then $\ell'(u_1) \in W_i, \ell'(u_2) \in W_{i'}$ for some $i,i'\in [n], i\neq
i'$ with $(v_i, v_{i'})\in E(G)$ (from property 5 of
Lemma~\ref{lem:reduction}).  Therefore $c(v_i) \neq c(v_{i'})$ so for $\alpha
\in [3]$, we have $H,\ell',\myc' \models \left(u_1 \in D_\alpha \to \neg (u_2 \in
D_\alpha)\right)$.

For the converse direction, suppose that $H,\ell,\myc \models \phi_{Col}$ for
the nowhere defined $\ell,\myc$. Then there exists a coloring function $\myc'$
such that $\myc'(D_\alpha) = S_\alpha$, for $\alpha \in [3]$ and $H,\ell, \myc'
\models \phi_{Col}[X_1\setminus D_1][X_2\setminus D_2][X_3\setminus D_3]$. We
extract a coloring of $V(G)$ as follows: for $i\in [n]$ we set $c(v_i)$ to be
the minimum $\alpha$ such that $W_i \cap S_\alpha \neq \emptyset$. We show that
the coloring $c:V(G) \to [3]$ defined in this way is proper. Consider $i,i'\in
[n]$ such that $(v_i, v_{i'})\in E(G)$. Let $\ell'$ be a labeling function such
that $\ell'(u_1)\in W_i \cap S_{c(v_i)}$ and $\ell'(u_2) \in W_{i'} \cap
S_{c(v_{i'})}$. Observe that $W_i\cap S_{c(v_i)}\neq \emptyset$ by the
definition of $c(v_i)$. Then $H,\ell',\myc'\models \phi_{adj}[x_1\setminus
u_1][x_2\setminus u_2]$.  Therefore we have that for $\alpha \in [3]$,
$H,\ell',\myc' \models \left(u_1\in D_\alpha \to \neg (u_2 \in
D_\alpha)\right)$.  Therefore $S_{c(v_i)} \neq S_{c(v_{i'})}$, which means that
$c(v_i) \neq c(v_{i'})$.  \end{proof}

\section{Conclusions}

We have given tight upper and lower bounds on the complexity of model checking
first-order and monadic second-order logic formulas parameterized by the vertex
integrity of the input graph. Our results are of course only of theoretical
interest, as the algorithms of \autoref{thmFOalg} and \autoref{thmMSOalg} are not
meant to be implemented in practice. One interesting avenue for further
research would be to extend our results to monadic second-order logic with
edge-set quantifiers, also known as MSO$_2$ logic. In the case of meta-theorems
for vertex cover, the extension from MSO$_1$ to MSO$_2$ is not too complicated,
as in a graph with vertex cover $k$, every set of edges can be described as the
union of $k$ sets of vertices (every edge is incident on a vertex of the vertex
cover, so it suffices to give, for each such vertex, the set of second
endpoints of the edges selected incident to this vertex). It would be
interesting to see if this basic argument can be extended to vertex integrity,
and whether this makes the complexity of model checking MSO$_2$ formulas
significantly worse than the complexity we gave for MSO$_1$ formulas.

\bibliographystyle{alphaurl}

\bibliography{vimeta}

\end{document}